\documentstyle[11pt,aasms4]{article}
\tighten
\slugcomment{Draft printed: ~~\today}
\begin{document}
% Needed TeX macros
\def\COBE{{\sl COBE\/}}
\def\wisk#1{\ifmmode{#1}\else{$#1$}\fi}
\def\um     {\wisk{{\rm \mu m}}}
\def\etal   {et~al.~}
\def\deg    {\wisk{^\circ}}
\def\icm    {\wisk{\rm cm^{-1}}}
\hfuzz=10pt \overfullrule=0pt
\pretolerance=10000
\raggedright
\pretolerance=1000	% hyphenation on

\title{Calibrator Design for the COBE 
\footnote{
The National Aeronautics and Space Administration/Goddard Space Flight
Center (NASA/GSFC) was responsible for the design, development, and 
operation of the Cosmic Background Explorer ({\sl COBE}).
GSFC was also responsible for the development of the analysis software and
for the production of the mission data sets.  The {\sl COBE} program was 
supported by the Astrophysics Division of NASA's Office of Space Science 
and Applications.
}
Far Infrared Absolute Spectrophotometer ({\sl FIRAS})}

\author{J.~C.~Mather\altaffilmark{2}, D.~J.~Fixsen\altaffilmark{3},
R.~A.~Shafer\altaffilmark{2}, C. Mosier\altaffilmark{2}, and 
D.~T.~Wilkinson\altaffilmark{4}}

\altaffiltext{2}{NASA/Goddard Space Flight Center}
\altaffiltext{3}{Raytheon STX}
\altaffiltext{4}{Princeton University}

\begin{abstract}
The photometric errors of the external calibrator for the FIRAS
instrument on the COBE are smaller than the measurement errors on the cosmic
microwave background (CMBR) spectrum (typically 0.02 MJy/sr, 1
$\sigma$), and smaller than 0.01\% of the peak
brightness of the CMBR. The calibrator is a re-entrant cone, shaped like
a trumpet mute, made of Eccosorb iron-loaded epoxy.  It fills the entire
beam of the instrument and is the source of its accuracy.  Its known
errors are caused by reflections, temperature gradients, and leakage
through the material and around the edge.  Estimates and limits are 
given for all known error sources. Improvements in understanding the 
temperature measurements of the calibrator allow an improved CMBR temperature
determination of 2.725$\pm$0.002 K.
\end{abstract}


\section{Introduction}
 
We describe and analyze the performance of the blackbody calibrator for the 
Far Infrared Absolute Spectrophotometer (FIRAS) instrument on the Cosmic 
Background Explorer (COBE) satellite. The COBE (Mather 1982, 1987, 1993, 
Boggess \etal 1992) was launched on November 18, 1989 on a Delta rocket, and 
carried three instruments to measure the diffuse infrared and microwave 
background radiation.  The primary goal of the FIRAS (Mather \etal 1993) was 
to compare the CMBR spectrum to a blackbody spectrum, the predicted ideal
result of a hot Big Bang.  Even small deviations from the blackbody form
would be important to cosmology, but none have been found. 

FIRAS results include limits on the distortion of the CMBR spectrum 
(Mather \etal 1994, Fixsen \etal 1996), the interpretation of these limits 
(Wright \etal 1994), measurements of the line emission of the Galaxy
(Bennett \etal 1994), a measurement of the CMBR dipole caused by
the Sun's motion (Fixsen \etal 1996) and a measurement of the 
spectrum of the intrinsic anisotropy (Fixsen \etal 1997).  The process of 
calibrating the instrument is described by Fixsen \etal (1994). The data
are available from NSSDC (National Space Science Data Center 
http://www.gsfc.nasa.gov/astro/cobe) and are described extensively in the 
FIRAS Explanatory Supplement (Brodd \etal 1997). The weighted rms
deviation between the CMBR and the calibrator blackbody is only 0.005\%
of the peak brightness, over the frequency range from 2 to 20 \icm\ (5
to 0.5 mm wavelength). The transfer of this result to an absolute
statement about the CMBR spectrum depends on the accuracy of the
blackbody calibrator, which is the subject of this paper.  The measured
deviations are an order of magnitude smaller than the accuracy that was
originally specified in the instrument design, so it is appropriate to
review the main sources of calibrator errors. 
 
The FIRAS measurements span the frequency range from $\nu$ = 1 to 97 \icm, in 
two spectral bands divided at 20 \icm.  It has a spectral resolution 
$\Delta \nu/\nu \sim 0.0035$ (FWHM) limited by beam divergence, and an 
unapodized $\Delta \nu \sim $ 0.4 \icm\ limited by the maximum path difference 
of 1.2 cm (0.09 \icm\ and 5.6 cm for the long or high resolution data). The
spectral resolution is obtained with a Fourier transform spectrometer
based on the Martin-Puplett (1970) polarizing form of the Michelson
interferometer.  It is fully symmetrical, with separation of
the two input and two output ports (Fig 1).  The inputs are coupled to
a reference and the sky or external calibrator.  All wavelengths are measured
simultaneously with the same detectors and the same optical path. 

The radiation accepted from the sky comes from a circle 7\deg\ in diameter, 
defined by a Winston cone (Welford \& Winston 1978, Mather 1981, 
Miller \etal 1982), also designated the Sky Horn.  The Sky Horn concentrates 
the beam into a circular aperture $2\surd{1.5} /\pi \simeq$ 0.78 cm in diameter.
The Sky Horn is a non-imaging device which scrambles the radiation paths.
The Sky Horn and instrument transmit an \'etendue of $A\Omega=1.5$
cm$^2$sr. The number of independent geometrical modes of the diffracted 
radiation field is $n=2 A\Omega \nu^2$, where the factor of 2 allows for
polarization states, permitting operation at frequencies as low as 1 \icm. 

Radiation passing through this aperture is focused into the spectrometer by 
a similar elliptical concentrating cone.  Radiation not passing through the 
aperture is returned to the sky or calibrator, except for a fraction lost by 
absorption in the horn walls.  The absorption of the Sky Horn was measured in 
flight by changing the concentrator temperature to determine the emissivity of 
the combined parabolic and elliptic
cones.  The horn emissivity is approximately $0.012 + 0.0015~$cm$ \times \nu$
in the low frequency channel (2 to 20 \icm).  By Kirchhoff's law, the
emissivity $\epsilon$ and the reflectivity $r$ are related by $\epsilon
+ r = 1$.  The calibrator and horn form a cavity in which most of the
radiation incident on the surface of the calibrator was originally
emitted by the calibrator, and has the same temperature. This fact
relaxes the requirement on the reflectivity of the calibrator by more
than an order of magnitude. 

The input end of the Winston cone is connected to a flared section like 
a trumpet bell, which suppresses diffracted sidelobes over a wide spectral 
band.  Sidelobe measurements have been reported (Mather \etal 1986), and are 
in good agreement with the calculations based on the geometrical theory of 
diffraction (Levy and Keller 1959).  The smooth
transition to a curved flare also suppresses diffraction at the
aperture, which would otherwise enable the instrument to see itself
through diffracted backscatter. 

The entire instrument operated in a vacuum, cooled to 1.5 K by
conduction to a superfluid liquid helium tank. A large external
conical shield protected the cryostat and instruments from direct
radiation from the Sun and the Earth. The Sun never illuminated the
instruments or cryostat, but the COBE orbit inclination combined with
the inclination of the Earth's equator to the ecliptic allowed the
Earth limb to rise a few degrees above the plane of the instrument and
sunshade apertures during about 1/6 of the orbit for 1/4 of the year.
During this period, the sky horn could not be cooled to
2.7 K because of the Earth limb heating. The edge of the shield is
approximately coplanar with the entrance aperture of the FIRAS
instrument, so there was no line of sight path for radiation from the
shield into the instruments.  The calibrator and its support arm project
above the aperture plane, and were exposed to radiation from the
warm parts of the cryostat and shade.  The effects of this radiation were
reduced by multilayer insulation and are estimated below.  The 
temperatures reached during the illumination by the Earth limb are 
described by Mosier (1991).

The other input (the reference input) also has a Winston cone, although
a smaller one, and a calibrator (the internal calibrator or ICAL, Fig 2). 
The length to diameter ratio for this reference cone (Ref Horn) is similar
to that of the main input cone (Sky Horn) to match the 
emission properties. The ICAL is mounted in the Ref Horn and could not
be moved. It is similar in some respects to the main calibrator (XCAL).
The ICAL and Ref Horn are useful in that they provide a signal much like 
that of the sky and Sky Horn. Because the interferometer measures the 
difference between the two inputs, this reduces the signal amplitude and 
relaxes the gain stability and dynamic range requirements of the FIRAS 
instrument by about a factor of 100.

The true comparison is between the sky and the XCAL, which provides the 
absolute reference by radiating into the same place with the same temperature
(at different times) with all of the other parts of the instrument in similar 
states. The ICAL has ~4\% reflections and possible gradients of several mK,
but the real requirement on it is that it is repeatable.  This was tested over
10 months on sky data as well as calibration data.

There are 3 requirements for the main calibrator.
1)It must have a well defined temperature which is known. 
2)It must have low reflectivity (emissivity = absorptivity = 1)
3)It must completely fill the beam (no leakage).

\section{Calibrator Design and Material} 

The main calibrator could be moved into the aperature on command. It was 
inserted for 3 days per month, and 3 days per week for the last 7 weeks.  
Its temperature was controlled by a servo loop using an electrical
heater and a germanium resistance thermometer (GRT).  The control range was
from 2 to 25 K, and the temperature was stable to within the 
GRT resolution of about 0.2 mK $(T/2.7~$K$)^3$.  The
temperature was monitored by 3 additional GRTs
in two separate self calibrating AC ohmmeter circuits. 
When not in use, the calibrator was kept in a protected well with the
active surface facing the sky. It was moved by a geared stepper motor.
 
The calibrator is illustrated in Figure 2. It is 140 mm in diameter, 
$\sim230$ mm long and shaped like a trumpet mute, with a central peak and a 
single groove, each with a full angle of $\psi$ = 25\deg. This shape was chosen 
to suppress specular reflections from the surface.  For specular 
reflections, a ray incident on the calibrator parallel to the axis 
(ie visable to the interferometer) must be 
reflected from the surface 7 times before it leaves the calibrator region.  

The calibrator was machined from two castings of Eccosorb CR-110 (Emerson and 
Cuming 1980), one for the central peak and one for the remainder, which were 
glued together using Eccosorb. Eccosorb is an epoxy loaded with
fine iron powder ($\sim5~$\um), with an admixture of Cab-o-sil, a
fine silica powder ($\sim10~$\um).  The silica powder makes the liquid epoxy
thixotropic, so that the iron powder does not settle during the curing
process.  

The optical properties of the Eccosorb have been reported (Hemmati \etal 1985,
Peterson and Richards 1984, Halpern \etal 1986). The normal incidence surface 
power reflection is approximately 0.08 + (0.06 \icm)/$\nu$, corresponding 
to a refractive index of about 2.  For the purposes of this paper we 
use a refractive index that produces the measured normal incidence 
reflection.  At low temperatures the absorption coefficient was found to
be  $\alpha \approx$ 0.3 \icm + 0.45 $\nu$ over the range of frequencies used
here. At 295~K the absorbtion coefficient is about 
twice as large. The manufacturer's literature shows that the 
permeability approaches unity at high frequencies, and we have assumed it is 1.
 
To achieve more rapid thermal equilibration, the back surface of the 
calibrator was covered with 0.25 mm copper sheets. 
Differential contraction prevents good adhesion, so the copper was 
perforated to allow the Eccosorb to penetrate it.  The copper was 
corrugated by geared rollers and is cut in the direction perpendicular 
to the corrugations to make it flexible.

An aluminum foil cap was placed over the back of the copper, since
the Eccosorb is not entirely opaque and there are gaps in the coverage
of the copper sheets. The back of this structure was covered with a
multilayer insulation blanket, containing 20 layers of aluminized
Kapton separated by layers of Dacron net.  This insulation was required
because a portion of the calibrator back was exposed to infrared emission
from warm portions of the spacecraft, from the Moon, and occasionally
from the Earth limb.

The calibrator was designed to have no steady state heat flow through
the absorber material, and therefore no temperature gradient.  The
copper backing was soldered to a copper ring, and the copper ring was
attached to the support arm.  The temperature of the ring was controlled
by the servo, with the electrical heater and the sensor mounted on the
ring.  This ideal concept was violated in potentially important ways.
The copper ring was not mechanically strong, so mounting bolts pass
through it into the body of the absorbing material.  Also, the manganin
thermometer and heater wires are thermal conductors, and carry some
heat away from the calibrator and down the support arm.  The heat
radiated into the horn by the calibrator is negligible because of the low
temperature, and most of it is reflected back to the calibrator.  A
thermal contact also exists at the calibrator edge.  The
calibrator did not touch the antenna, leaving a gap of 0.6 mm which was
spanned by two ranks of aluminized flexible Kapton leaves 0.1 mm thick
and about 12 mm long. The contact force is small and the estimated thermal 
contact is small as well.  Moreover, under the most important calibration 
conditions, the antenna was kept at the calibrator temperature, 
guaranteeing negligible heat flow.
 
The temperature of various components of the FIRAS were monitored with
GRTs. The GRTs were measured with sine 
wave excitation at two current levels (0.4 and 6.4 $\mu$A).  The thermometers 
have separate current excitation and voltage leads, so the lead resistances
in the cryostat had little effect on the results. The excitation
frequency was 40 Hz, low enough to minimize shunt capacitance effects.
The FIRAS has two separate ohmmeters, each used to read its own set of
thermometers.  Each ohmmeter circuit includes a stabilized sine wave
oscillator and current source, a wideband amplifier, a phase sensitive
detector, and a 14 bit analog-to-digital converter.  The input to each
ohmmeter is multiplexed through a MOSFET switch, so that it can read 16 
thermometers and 4 calibration resistors, some with both low and
high currents, in every major frame of the telemetry (32 sec).

The temperature was controlled by a servo loop that used a separate GRT.
The servo provided a choice of thermometer bias currents of 1, 4, 16, and 64 
$\mu$A, to account for the wide range of thermometer resistance over the 
temperature range from 2 to 20 K ($\sim$7000 to 45 $\Omega$). It also 
provided adjustable gain factors of $2^n$ for $n=0...7$
for both proportional and integral gain.  For low temperatures around
2.75 K, the dominant time constant was only 16 sec, but at 20 K it was
about 14 min.  The square law nonlinearity of the heater
made control of positive temperature steps difficult.  The necessary
heating power was approximately $(T^2 -T_{\hbox{min}}^2)*100\mu$W/K$^2$, 
where $T_{\hbox{min}}$ was the minimum temperature (about 2.2 K in flight) 
in the absence of heater power. 

\section{Error Analysis}

There are many possible errors at a level of a few parts in $10^5$.  The
temperature of the Eccosorb must be measured and uniform through the
thickness of the material and across the aperture. The reflectance must
be small, so that other objects illuminating the calibrator do not
contribute significantly to its output.  There must be a good seal
around the edge, so that radiation from outside does not leak into the
beam.  Each of these error sources is discussed below.  

\subsection{Absolute Temperature Scale}

The exact temperature of the CMBR is not important for cosmology, since
every other cosmological constant is more poorly determined.
However, spectrum distortions are important and require the comparison
of the results of different instruments.  The FIRAS measurement
for $T_{\hbox{cmbr}}$ was 2.728 $\pm$ 0.004 K as described by Fixsen \etal (1996).  
The uncertainty was entirely due to estimates of systematic errors. In 
particular a discrepancy of 4.5 mK between two methods of determining the
temperature scale led to the uncertainty estimates. This discrepancy has been
resolved allowing a better determination of the CMBR temperature.

The calibrator has three GRTs, two attached to the copper heater ring and 
one embedded in the Eccosorb tip (see Fig 2). They disagree by 3 mK, significantly greater than 
the expected precision of 1 mK.  The thermometers were calibrated against 
a standard thermometer from the National Institute of Standards and
Technology, and their calibration was better than 1 mK at the time.
Relative to their mean, the three show deviations of -3.3, -.3, and +3.6 mK at
2.7 K. Based on only two degrees of freedom, this absolute temperature scale 
has a 1$\sigma$ uncertainty of 2 mK. At higher temperatures the GRT at the tip 
of the cone deviated more from those at the copper ring. The tip GRT was not
used in the final calibration.

The ICAL had 2 GRTs in addition to the GRTs used by the temperature control
circuitry.  Here too, the tip GRT read warmer than the GRT at the base.
An additional drift of $\sim3$ mK was noted in the early part of the mission.
The drift was more pronounced in the tip GRT than the base GRT.

To further investigate possible drift in the GRT
calibration, a group of 10 thermometers was recalibrated 1.7 years after 
the launch, and while 7 remained within 1 mK of their original response, 3
deviated by as much as 6 mK. Some of the 
recalibrated thermometers were more susceptible to self-heating by the 
excitation current, a temperature shift 
proportional to the square of the excitation current. The hermetically 
sealed helium filling, which helps establish thermal contact, might have 
leaked out. 

The calibrations were made with a 1 Hz square wave excitation, while the 
flight ohmmeter used 40 Hz sine waves. Direct comparison of the calibration
system and the flight system 
showed an offset of about 7 mK. The explanation of this was not determined.
In the comparison experiment neither system was in its final configuration
and either (or both) of them could have been affected by the requirement
of long cables to the dewar that were not used in flight. Nonlinearities of 
the flight ohmmeter were measured with the calibration resistors in flight, 
and would not cause an error larger than 0.3 mK at 2.7 K.

The flight ohmmeters used two different excitation currents. These
typically had differences of 5 mK with the higher excitation current reading
a higher temperature both in ground tests before the flight and in 
the flight. The higher excitation current was used at 2.7 K because the 
lower excitation current had more noise. Averaging over ~100000 samples
allows comparison at the 10~$\mu$K level. The offset is not uniform and
varies between 2.5 mK and 7.5 mK on the XCAL thermometers. 

The self-heating power at 2.7 K is 110 nW for the high current and 0.4 nW
for the low current setting. A temperature change of 5 mK implies a 
thermal conductivity of 22 $\mu$W/K.  At higher temperatures the
heating is smaller and the thermal conductivity is larger, thus one
is led to expect the self-heating to be only 10\% as large at 5 K.
This general trend can be seen in the data but the details are obscured
by noise (although there are $\sim 100000$ observations at 2.7 K there
are only a few thousand between 2.8 and 5.5 K). Although it is possible to 
determine the difference between the low and high current readings to 
10 $\mu$K the ultimate accuracy is no better than 1~mK. This correction
of 5 mK leads to a new CBR temperature estimate of 2.725 K rather than
the 2.730 K previously reported for the thermometers.

A 5 mK error in the temperature determination of the XCAL leads directly
to a 5 mK error in the temperature determination of the CMBR.  However,
the calibration process corrects other effects of the error to first order
(Fixsen \etal 1994).

The FIRAS also allows other determinations of absolute temperatures,
based on the wavelength scale and the known shape of the Planck
function. The detector noise contribution to the uncertainty of this
scale is quite negligible, and the largest known uncertainty is the
determination of the wavelength scale.  It is derived from FIRAS observations 
of the interstellar CO and [C~I] lines at 1300, 867, 650, and 609 \um\ 
(Fixsen \etal 1996). The temperature scale was determined independently from 
7 different combinations of the four detectors and four scan modes. These 
determinations agreed within their uncertainties and the weighted uncertainty
is 0.2 mK. There is an additional common uncertainty of 0.82 mK due to the 
uncertainty of the frequency scale.  The result is 2.7255~K~$\pm$.85~mK.
With the correction for self-heating the discrepancy with the GRT measurement
is only 0.5 mK, within the uncertainty estimates of either method.

There is yet another determination of the temperature which is independent of
the previous two. The FIRAS measured the dipole amplitude, 3.372 $\pm$
0.007 mK (Fixsen \etal 1996) and the shape of this spectrum was fit to a 
$\partial B/\partial T$ with an adjustable temperature, with the result of 
2.717 K. The DMR on board COBE also measured the dipole amplitude, 3.353 
$\pm$ 0.024 mK (Bennett \etal 1996), and the DMR was calibrated independently 
to 0.5\% (Bennett \etal 1992). The DMR calibration was also checked by 
measuring the dipole effect of the Earth's motion around the Sun.  
By assuming the DMR dipole amplitude is correct and the discrepancy is due to
a calibration error we can correct the FIRAS by dividing the frequency
scale by 1.002, which multiplies the dipole temperature by the same factor.
The final result is 2.722~K~$\pm$12~mK which is $3\pm 12$ mK below the final
temperature scale. The uncertainty is dominated by the uncertainty in the 
DMR calibration.

Averaging these three determinations of the CMBR temperature 
$<2725 \pm 1, 2725.5 \pm 0.85, 2722 \pm 12>=2725.28 \pm 0.66 $ mK.
The $\chi^2$ is 0.29 for 2 degrees of freedom. There is reason to be cautious 
but a CMBR temperature of $2.725 \pm 0.002$ K (95\% confidence) is a good 
description of the CMBR.

\subsection{Thermal Gradients}
 
The main thermal requirement for the FIRAS calibrator is uniformity.
By design, the calibrator has no steady state heat flows through the 
absorbing material, so that theoretically there can be no gradients in 
temperature.  The absorber is supported on a copper ring (140 mm dia, 
13 mm wide and 3 mm thick) whose temperature is regulated by a heater, and 
heat from the copper ring flows out directly to the helium tank through a 
strap.  However, compromises were required to make the calibrator survive 
launch vibrations, and the lead wires to the thermometers carry a small amount 
of heat away from the calibrator.  To investigate the effects of these 
compromises, an accurate copy of the flight calibrator (the flight spare) was 
built and instrumented with thermometers and heaters.  In addition, a finite 
element numerical model was devised and adjusted to match the laboratory test 
data. We report first on the test data, then on the numerical model, and 
finally give the estimated uncertainties induced by the gradients.

\subsubsection{Thermal Gradient Test}

The calibrator copy was instrumented with 4 additional
thermometers, as illustrated in Figure 3, and mounted in a helium
cryostat suspended by threads in a vacuum.  A copper heat strap with a conductance
similar to that of the flight calibrator mounting arm and cooling strap
connected the calibrator mounting ring to the helium bath.  An 
aluminum shield prevented radiation from warm parts of the cryostat from
reaching the calibrator. The thermometers were from the same group as
the flight thermometers and were recalibrated before installation (see sec 3.1).

Temperature gradients were measured for a range of helium bath and calibrator 
temperatures. Residual thermometer calibration problems were still present, at 
the level of $\sim 1$ mK. They were recognized by turning off the calibrator 
heater, so that there should be no genuine temperature gradients, and 
adjusting the helium bath temperature to 2.634 K by regulating the helium 
pressure.  The measured calibrator temperature was 2.694 K, confirming that the 
residual heating sources such as radiation from the warm sections of the 
cryostat were negligible. Then the helium temperature was lowered to 1.5 K 
while the temperature control servo was activated to fix the calibrator
temperature at 2.7 K.  The change in measured calibrator
temperatures that occurred as the helium bath was cooled are considered
to indicate real thermal gradients.  Those thermometers mounted
together on the copper ring with the heater and control thermometers
showed temperature changes of less than 1.3 mK.  Those mounted near the
tip of the calibrator deep inside changed less than 0.1 mK.  Some
thermometers were also glued to the surface of the cone that faces the
spectrometer. These changed less than 1 mK.

\subsubsection{Thermal Gradient Model}

These measurements confirm that the temperature gradients within the
calibrator material are small but detectable under some circumstances.  
To understand their origin, and to estimate their values in the flight 
calibrator, we made a finite element numerical model of the temperatures, 
as illustrated in Figure 3. Eccosorb has a thermal conductivity of 0.8 mW/cm K
(Halpern 1986), while the copper has a conductivity of 2 W/cm K at 3 K.  
The corrugated copper backing material is 0.25 mm thick, but
its lateral conductivity still exceeds that of the thick Eccosorb by a
factor of 3.  The boundary impedance between the copper and the
Eccosorb is not known and may be relatively high if the
adhesion between them is broken by the differential thermal
contraction. The heat capacity of the Eccosorb is  $C_p=0.6 T^{2.05}$ mJ/gK 
(Peterson and Richards 1984  citing a private communication from M. Halpern), 
and its mass $\sim1.5$ kg. The slowest time constant is for the
thermometer attached to the exterior of the Eccosorb, and is 16 sec at 2.7
K.  This implies that the copper-to-Eccosorb conductance is greater than
.35 W/K and the contact area is $\sim 100$ cm$^2$. 

There is one significant discrepancy between the model and the actual
calibrator.  The time constant for the thermometers on the heater ring to
sense the change in applied heat is 21 sec, which is much longer than
the expected value. This causes considerable difficulty in tuning the control
servo, since the time delay causes phase shifts that limit the servo
gain and hence its speed of response.  The time constant could be
important to the thermometer accuracy if it indicates that the
thermometer is not well attached. In that
case its lead wires would conduct heat from the thermometer down to the
helium bath, and the thermometer would read too low.  This effect was
checked in the calibrator copy and would have been detected unless all
the thermometers showed exactly the same amount of error, an unlikely
coincidence.  It is more likely that the heater itself is not well 
attached to the ring, which could happen if its adhesive failed at low 
temperatures. To guard against this failure, a spring plate was
added to press the heater against the copper ring in both the flight and
copy calibrators. The heater is a Minco resistive film embedded in a 
Kapton insulating sandwich, and provides heat distributed evenly 
around the copper ring.  The adhesion could not be verified in
the cooled calibrator because the adhesive becomes sticky again at room
temperature. 

\subsubsection{Thermal Gradient Error}

The FIRAS beam may see different parts of the calibrator and penetrate to
different depths at different wavelengths.  We consider three possibilities,
and based on the tests and numerical model, we conclude that these 
effects are small. To illustrate, we consider example calculations.

First, consider a radial gradient in calibrator temperature.  Our 
measurements limit such a gradient to $\sim 1$ mK.  To obtain a first order
spectrum error we require an antenna pattern that is dependent on both
radius and frequency.  In geometrical optics there is no reason for such a
dependence, and indeed the concentrator antenna is a good scrambler.  A 
plausible guess would be that wave effects could cause the low frequency beam 
to avoid the walls of the concentrator.  Without a detailed calculation 
we can still assume a Taylor series expansion in wavelength, but we do 
not know whether the leading term is linear or higher order. 
Taking the maximum error as 1 mK at the cutoff wavelength of the
antenna $\lambda_0^2 = A \Omega$, we plot the resulting photometric
error in Figure 4 for an assumed linear leading term.

Second, suppose that there is a cold spot in the calibrator that is seen 
through the thickness of the Eccosorb (12 mm). Such an error could occur if 
the mounting bolts induce 
a gradient.  Figure 4 shows the error resulting from a spot of 
area 0.5 cm$^2$ at a temperature 100 mK different from the average. The 
attenuation coefficient of the Eccosorb is strongly frequency dependent,
so that the resulting error falls rapidly with frequency.  

Third, suppose that there is a temperature gradient with
depth into the calibrator material.  We have no reason to expect this to
be a dominant term, since there is no radiative or other heat transport across 
the surface. Nevertheless, we can evaluate the effect as a
function of frequency.  The effective temperature is measured at an 
optical depth of unity into the Eccosorb, or a physical depth of 
$1/\alpha$.  Assuming that the gradient is 1 mK/cm, we obtain the error 
plotted in Figure 4.

There is a negligible second order effect due to possible
gradients. The Planck function for the mean temperature is not the same
as the mean of the Planck functions for the various parts, because it
is not linear in temperature.  This effect can be modeled precisely. 
The Planck function $B_{\nu}(T)$ can be expanded in a Taylor series
around the mean temperature, giving an effective radiated power

\begin{equation} 
     P_{\nu} = B_{\nu}(T) + \Delta T {\partial B_{\nu}\over 
\partial T} + {1\over 2} {\Delta T ^2}{\partial^2B_{\nu}\over\partial T^2},
\end{equation}
  
where $T$ is the original temperature, $\Delta T$ is the mean shift in 
temperature, and $\Delta T^2$ is the variance of the temperature
distribution.  We define $x = h \nu c/k_BT$, where $h$ is Planck's
constant, $\nu$ is the frequency $1/\lambda$, and $k_B$ is
Boltzmann's constant. The first and second derivatives are 

\begin{equation}
{\partial B_{\nu}\over \partial T} = B_{\nu}{x\over 
T}{e^x\over{e^x-1}},
\end{equation}

\begin{equation} 
     {\partial^2B_{\nu}\over \partial T^2}={1\over T}
{\partial B_{\nu}\over \partial T} 
	{\Bigl(x{{1+e^{-x}}\over{1-e^{-x}}} - 2\Bigr)}.
\end{equation}

This is closely related to the cosmological Compton distortion given by 
Sunyaev and Zel\'dovich (1969), scaled by the parameter $y$. A linearized 
form for the distortion of the spectrum  $S_{\nu}$ is 

\begin{equation}
{{\partial S_y}\over{\partial y}} = T^2  {\partial^2B_{\nu}\over \partial T^2}
-2 T {\partial B_{\nu}\over \partial T} 
\end{equation}

\begin{equation}
y = \int {{k(T_e-T_\gamma)}\over{m_ec^2}} d\tau_e\quad,
\end{equation}

where $T_e$, $T_\gamma$ and $\tau_e$ are the electron temperature, the  CMBR
photon temperature, and the optical depth to electron Compton scattering. 
This form preserves the number of photons in the spectrum.  Note that 
any term proportional to ${\partial B_{\nu}/ \partial T} $ 
is equivalent to a shift in the mean temperature of the CMBR, and still 
represents a pure blackbody spectrum.

The results found by Fixsen \etal (1996) from the FIRAS data are 
$y= (-1 \pm 7) \times 10^{-6}$. The corresponding equivalent
range of the cosmic temperature distribution is found from
$2y=var(T)/T^2$, where $var()$ is the variance. If we assume that the 
true CMBR spectrum has $y=0$, we conclude that the rms
variation of the calibrator temperature is therefore less than 15 mK.  
The limitation from the thermometers is tighter so any real gradient
must be insignificant photometrically. 

\subsection{Calibrator Reflectance}

The calibrator is the dominant radiator in a cavity bounded by 
four surfaces: the calibrator, the compound concentrator horn, the 
small aperture of the horn that leads to the spectrometer, and the 
gap between calibrator and concentrator.  If all these surfaces were at 
the temperature of the sky, and the sky were a blackbody, then the radiation 
inside the cavity would be perfect blackbody radiation and would 
have the same intensity and spectrum as the sky radiation.  In that 
case, moving the calibrator in or out of the beam would make no change 
of the radiation field.  This is the basis for a precise differential 
comparison of the sky to the blackbody.

The leading deviations from the perfection of this blackbody cavity are 
as follows.  First, the transmission of the horn for the emission from 
the calibrator is not unity, but the transmission is absorbed in the gain 
constants of the spectrometer by the calibration algorithm.  Second, the 
horn is not always at the same temperature as the sky or calibrator, but this 
is also included in the instrument model and the horn emission is measured in 
the calibration process. The calibration process also includes terms for 
emission from the dihedral, the collimating mirrors, and the bolometer itself.
The leading term that is not included in the calibration model is the change
in the emission of the instrument, at a temperature of about 1.5 K, that is
induced by inserting the external calibrator. This term can not be included 
in the calibration model because the calibrator must be inserted to calibrate and must
be out of the horn to view the sky. To first order, the error introduced is 

\begin{equation}
dP = (B(T_i) - B(T_c))\quad r_i
\end{equation}

where $dP$ is the error in emitted intensity, $B$ is the Planck function,
$T_i$ and $T_c$ are the instrument and calibrator temperatures, and
$r_i$ is the reflectivity of the calibrator for radiation
originating in the instrument and reflected back toward it.

\subsubsection{Reflectance Measurement}

Direct measurements of the calibrator reflectance in the same geometrical 
configuration used in flight were made, but at room temperature.  The 
refractive index of the Eccosorb
is nearly the same at room temperature and cold, so the diffraction and
surface reflection should be the same.  A coherent microwave system
was built to illuminate a duplicate calibrator through a duplicate of
the flight antenna.  Measurements were made from 30 to 37 GHz and at
93.6 GHz (Fig 5).  Radiation from a source passes
through an attenuator and frequency meter to a ``magic tee" microwave
beamsplitter. From the tee, radiation splits between two arms of the magic tee.
One arm is terminated with a load and the other has
waveguide leading to the small end of the FIRAS antenna. A
rectangular-to-circular transition couples the guide to the
antenna at the 0.78 cm diameter throat aperture.  The radiation is
collimated by the antenna and is incident on the calibrator in its usual
position.  Radiation reflected from the calibrator returns along its
path and some of it is split off by the ``magic tee" to the detector.  At
that point, it interferes coherently with the other signals already
present.  The total intensity reaching the detector depends on the phase
and amplitude of the radiation reflected from the calibrator. 

The calibrator was moved along the axis of the antenna to vary the phase
of its reflection, that interferes with the larger reflections from the joints
and other parts of the setup, producing a sinusoidal interference pattern that
clearly identifies the part due to the calibrator reflection.  The
calibration of the method was made by substituting a flat metal plate
for the calibrator.  The amplitude was also measured as a function of
the angle between the calibrator axis and the antenna axis (Fig 6).  The
response was greatest when the angle was zero, and there was no
pronounced structure between the measured points. 
The peak response was -55.8 $\pm$ 1.5 dB at 33.4 GHz and
-59.0 $\pm$ 1.5 dB at 93.6 GHz.  The frequency was also swept from 30 to
37 GHz to search for resonant enhancement effects. The response was
greater at 35.25 and 36.86 GHz, where the response was increased to
-45.8 $\pm$ 1.5 dB at zero tilt angle. 

Note that the test apparatus differs from the flight configuration in an
important way: the flight antenna receives $n=2 A \Omega \nu^2$ modes,
and the waveguide system receives only one.   For 33.4 and 93.6 GHz
these values are $n$= 3.7 and 29.  Also, the multimode concentrator illuminates
the calibrator over a range of angles, and accepts reflected radiation
for the same range.  Therefore the response pattern, which is obtained 
with a single mode system and a plane wave illumination, should be convolved
twice with the measured beam profile to obtain a suitable average.  The
computed effective power reflectance based on these data is $3\times 10^{-6}$
at 33.4 GHz and $9\times 10^{-6}$ at 93.6 GHz.  Unfortunately the
angular response pattern was not measured at 35.35 and 36.86 GHz, so to
be conservative we increase the estimated reflection at those
frequencies by a factor of 10, giving $3\times 10^{-5}$ for the
effective reflectance.  The interpreted measurements are all less than
$3\times10^{-5}\icm/\nu$, which as we shall see below is about as expected from
diffraction calculations.  They are also too small to be detected in the
FIRAS photometry. 

\subsubsection{Diffractive Reflectance}

The diffraction of radiation at the calibrator was modeled using the simple 
Huygens' Principle for scalar waves (Levy \& Keller 1959).  We assume that an 
incident plane wave comes up the concentrator horn, and calculate the field 
incident on the calibrator surface.  The undisturbed plane wave would 
not diffract back toward the source in a simple cylindrical pipe, so we 
use the reflected wave amplitude as the effective source for the
calculation.  To account for polarization effects at the first 
surface reflection we use the root mean square value for the two
amplitude reflection coefficients.  To account for the multiple 
reflections of the waves as they go into the groove, we calculate the 
sum of the amplitudes of all the specularly reflected waves, up to 7 
reflections. This enhancement is only a factor of 1.1 in amplitude, and is 
applied to the incident amplitude after multiplication by a Gaussian to 
connect it smoothly with the rest of the surface. 

The numerical integration was performed in cylindrical coordinates.  The 
integral with respect to angle yields a Bessel function of zero order.  
The integral with respect to radius was done numerically with 10000 
steps.  The results are plotted in figure 7.  They show a dependence on 
angle that resembles the Bessel functions, as expected, but do not show 
the same details as the experimental results.  Following the same
prescription as for the experimental data, we also computed the mean
reflection coefficients averaged over the solid angles of the source and
receiver antenna.  These are also plotted in figure 7, and are
approximated by $3\times10^{-5}\icm/\nu$.  There is reasonable agreement 
between the theory and the measurements, considering the number of 
approximations made in both. 

\subsubsection{Specular Surface Reflectance}

To estimate the specular reflectance of the calibrator, we
approximate it by a V groove of the same included angle in an
infinitely thick medium.  In this case, there is no mixing of
polarizations, and all the angles of incidence are known.  A ray
originating in the spectrometer will be reflected 7 times before
exiting the V groove, at angles from normal of +12.5, -37.5, +62.5,
-87.5, +112.5, -137.5, and +162.5 degrees where the + indicates away from the
center axis and - indicates towards the center axis.  We estimate the 
refractive index from the normal surface reflectance $R_s$ and use the
Fresnel formulas to compute all the reflectances.  Averaging
over polarizations gives $R_{\hbox{spec}} = 5 \times 10^{-5}$ at 1 \icm.
 
Only a fraction of the returned specular beam is directed back
toward the spectrometer.  The horn defines a circular field of
view of 7\deg\ diameter, so one may consider
that it sends a circular bundle of rays toward the V groove.
The circle comes back shifted over by 5\deg\
because of the accumulated effect of the 7 reflections, so that
it overlaps slightly with the circle representing the rays that
can be received by the instrument.  The fraction of the area
that overlaps is computed as 14\%. In other words, if the
calibrator were a V groove of two mirrors, 14\% of the beam
originating in the spectrometer would return to it to be
detected.  Since the actual calibrator is not a simple V groove,
the overlap fraction could be either smaller or larger, but the
computation is not precise enough to merit detailed attention. The final 
effective specular reflection is then $7\times 10^{-6}$, which is 
comparable to the diffractive term at about 4 \icm.
 
\subsubsection{Diffuse Surface Reflectance}
 
The surface texture is similar to that of a machined metal
surface having a surface roughness of $\sigma$ = 5 micrometers
rms.  We approximate the calibrator diffuse reflectance by
$R_{\hbox{surf}} = 4 R_n (\Omega/\pi) \sin  (\psi/2) (\sigma k)^2$,
where $k = 2\pi/\lambda$ is the wavevector, $R_n \approx 0.1$
is the normal reflectance of a polished surface, $\psi$ =
25\deg\ is the full angle of the cone and
groove, and the sine function accounts for the angle of
incidence of radiation from the spectrometer.  Evaluating at a
wavelength of 1 mm, we find $R_{\hbox{surf}} = 3.2 \times 10^{-7}$, quite
negligible, showing that a more exact calculation is
unnecessary.

\subsubsection{Internal Reflection}

The Eccosorb is not thick enough (12 mm) to be entirely opaque, so at cm 
wavelengths the back surface is partly visible through it.  The back surface 
is partly covered with irregularly shaped copper foils.  We take
its surface reflectance to be unity with a Lambertian angular distribution 
inside the Eccosorb.  
$     R_{\hbox{back}} = (\Omega/n^2\pi) T_s^2 \cos(\phi) e^{-2\alpha t}$,
where $T_s = 1 - R_s$ is the transmittance of the front surface,
$\alpha$ is the measured absorption coefficient of the material, and $t$
is the thickness.  The factor $1/n^2$ accounts for the change
in beam divergence at the refractive surface, and the $\cos(\phi)$ 
accounts for the spreading out of the radiation over the larger surface 
inside the material, with $\phi$ = 77.5\deg.  The surface transmittance is 
evaluated for an angle of incidence of 77.5\deg\ from the normal.  Evaluating 
this at $\nu$ = 1 \icm, we find $R_{\hbox{back}} =  2\times 10^{-5}$.  This number is 
already negligible at this frequency because of the relatively lower
sensitivity of the FIRAS, and decreases exponentially as the frequency
increases. 

The back surface reflections would be much more important if there were 
a specular glint within the material.  The Eccosorb shape is too complex 
to be amenable to easy calculation.  As an example to show the scale 
of the problem, consider a glint.  Because of the 
symmetry of the calibrator, a ray from the spectrometer can refract and
reflect all the way back to the instrument. The net reflectance of such 
a spot would be governed by a simple formula in the geometrical 
optics limit:
$     R_{\hbox{spot}} =  T_s^2 {r_{\hbox{spot}}^2/{r_{\hbox{cal}}}^2} 
e^{-2\alpha t}$.
If we assume that $r_{\hbox{spot}}$ = 0.3 cm is an effective spot radius, and
$r_{\hbox{cal}}$ = 7 cm is the calibrator radius, this gives $2.4\times 10^{-4}$ at 
1 \icm, which is not as small as other terms calculated above.  The estimates 
from this formula are plotted in figure 8.  
At room temperature, the attenuation coefficient of
the material is twice as large as at low temperatures, so in the
laboratory test such a spot would have given a reflectance of only
$3.5\times 10^{-5}$. This is comparable to the directly measured maximum
values in the 30 to 37 GHz band, and therefore might be a real
possibility. 

\subsection{Horn Emission}

Reflection of the horn emission by the calibrator would be significant
if the horn temperature were not controlled to match the calibrator
temperature.  Its emissivity is small but its area is large.  To first
order, it causes an error $dP_h = (B(T_h) - B(T_c)) \epsilon_h r_h$,
where $dP$ is the error in emitted intensity, $B$ is the Planck
function, $T_h$ and $T_c$ are the horn and calibrator temperatures,
$\epsilon_h$ is the horn emissivity toward the calibrator, and $r_h$
is the reflectivity of the calibrator for radiation originating in the
horn and reflected toward the instrument. We measured the horn
emissivity in the calibration process (Fixsen \etal 1994), and it is
small at low frequencies: $\epsilon_h \simeq 0.012 + 0.0015 (\nu/1$ \icm).  
We also set $T_h$ to match $T_c$ within 20 mK. The error
introduced here can be compared to that from the instrument radiation
described above: 

\begin{equation}
{dP_h\over dP} = {{B(T_h) - B(T_c)}\over {B(T_i) - T(T_c)} }
{{\epsilon_h {r_h}}\over{r_i}}, 
\end{equation}

Conservatively assuming that the ratio of reflectances is unity, we
still find that the error from horn emission and calibrator reflectance
is small relative to the error from instrument emission and
calibrator reflectance. 

The curious reader might also ask about the emission from the part 
of the antenna located above the calibrator.  This term is
negligible for three reasons: the emissivity of the surface is small,
the antenna quickly curves away from contact with the main beam, and
finally, the antenna is maintained at the temperature of the sky so that
its emissions compensate for the sky radiation that it absorbs.  To be
quantitative, we use the measured emissivity of the antenna.
A simple approximation for the horn emissivity
shows that it should be proportional to the surface resistance and the
effective length-to-diameter ratio $\int dx/D$, where $x$ is the
coordinate along the length and $D$ is the diameter.  Without doing a
proper calculation of the effects of the curvature, it is reasonable to
estimate that the effective length-to-diameter contribution of the horn
flare is less than 1\% of the total, so that its emissivity should be
less than a few parts in $10^4$. As stated above, its temperature is
also well known and kept within 10 mK of the sky temperature, and
therefore this error is less than a few parts in $10^6$.

\subsection{Edge Leakage} 

To prevent wedging the calibrator into the horn in flight, we required a
clearance between them.  The gap, about 0.06 cm,
could cause errors in the calibration.  In addition, there is a similar
area of the Eccosorb rim which could not be covered completely with
aluminum foil and multilayer insulation, and some radiation could be
transmitted through it.  The sources of warm radiation in flight
are the multilayer insulation blanket on the outside of the calibrator
and its support arm, the radiation emitted by the sunshield and
scattered downward by the arm, and the far infrared sky including the 
Moon, which shone onto the top of the calibrator for half of each 
calibration orbit. 

\subsubsection{Leakage Measurement}

We begin with the measurements and then make theoretical
extrapolations and interpretations.  There were three kinds of 
measurements, some taken on the ground with a warm cryostat dome above the 
instrument, the ordinary calibration data in flight, and finally some data
taken in flight with the calibrator moved progressively farther out of
the antenna.  

The ground data were taken with the FIRAS in the flight cryostat and
oriented so that the calibrator could be moved in and out step by step. 
First, the FIRAS observed the warm dome of the cryostat, at a
temperature near 40 K. The instrument calibration showed that the warm
dome had an effective emissivity of $\sim$10\%.  Second, the calibrator
was put in place, servo controlled to a fixed temperature around 13 K, 
and an hour of observations were taken. Then, the calibrator was
moved out in steps, with an hour of data at each position.  A single
step corresponds to about 1.4 mm of motion.  Data were taken for
positions 1, 3, 6, and 8 steps from the origin.  The broadband leakage in
the low frequency channel (1 to 20 \icm) was measured by the height of
the peak of the interferogram. Relative to the signal level seen
observing the dome directly, the peak height was approximately $(1.2 n^2
\pm 2)\times 10^{-6}$, where $n$ is the number of steps taken from the
nominal position.  Similar analysis of the high frequency channel (from
20 to 100 \icm) yielded a factor of $(3 n^2 \pm 35) \times 10^{-6}$.
Other observations were done with the calibrator at 4 K and while varying
the dome temperature from 5 to 66 K. These showed no evidence of leakage
around the calibrator.

We conclude that the attenuation is good when the calibrator is all the
way into the horn, and that it gets rapidly worse when it is pulled out
far enough for the flexible leaves to lose contact.  The evidence for
the quadratic dependence on $n$ is weak given the signal-to-noise ratio.
For further calculation we assume that the correct number is $4\times
10^{-6}$ when the calibrator is in place, in both low and high frequency
bands.  This is a reasonable number, corresponding to an illuminated gap
area which is 2\% of the total calibrator area, and an attenuation
factor of about 1.4\% for each row of flexible leaves. 

A test was also done in flight by removing the calibrator from the horn 
12 steps, or 17 mm.  Only a few interferograms were taken but there was 
no sign of a change of signal level.  Calculations based on the ground 
test data (see below) showed that it should be impossible to observe any
signal without spending a large fraction of the mission on this test, 
and in any case this test would not be relevant to the situation 
where the calibrator is in its proper position.  This test produces a 
weak limit, illustrated in figure 9.  The limits found from the flight
test were extrapolated from $n=12$ to $n=1$ according to the $n^2$ form
found in the ground tests, and are plotted as a horizontal line since no
spectral information was obtained.  While suggestive, this cannot be 
considered a firm limit.

The flight calibration data are sensitive to leakage at 
high frequencies, because most of the calibration data were taken with both 
calibrators and both antennas controlled to 2.7 K.  With temperatures 
this low, there is nothing in the instrument that can emit significantly 
above 30 \icm, so any signal seen there must be an instrument error.  
We used all the data taken in this configuration, and 
computed the residuals from the calibration model.  The spectrum of the 
residuals is nearly flat and noise limited except near 73 \icm\ where there are 
residual effects of the coherent instrument vibrations.  The weighted 
rms residual was 0.008 MJy/sr from 55 to 70 \icm. A limit can also be 
obtained by fitting an assumed spectrum to the residual.

\subsubsection{Leakage Interpretation}

To use these broadband limits and measurements we must assume the 
effective spectral shape of the radiation leakage.  The incident 
radiation comes from the multilayer insulation above the calibrator, 
from the sunshield and scattered from the calibrator support arm, from 
the Moon, and from the general far infrared sky. The spectra for the warm 
cryostat dome (measured on the ground) and for the Moon are reasonably 
accurate, but the multilayer insulation emission and scattered sunshield 
emission are order of magnitude estimates.

The blanket temperatures are quite uncertain but important.  They are
unmeasurable because the blankets are so thin and light that a thermometer
attachment would completely change their thermal properties.  The
insulation near the calibrator is exposed only to radiation from the sky
and should always be quite cold, but the insulation on the sides of the support
arm sees radiation from the interior of the sunshield at 180 K.  An 
unsupported gray body at this location would reach a temperature of 
about 40 K based on the view factor and the emissivity of the sunshield, 
assumed to be 0.05.  The thermal conductance between the blanket layers
and laterally to the corners where they are attached is sufficient to 
lower the temperatures to about 4 K, but this calculation is also 
quite uncertain and cannot be proven.  To be conservative we assume that 
the blanket temperature is 35 K.  The emission of the blanket towards 
the calibrator is assumed to have an emissivity of $0.003 \times 
\nu^{0.5}$ (approximately twice the calculated value for good aluminum at room 
temperature), and to fill 10\% of the solid angle at the top of the leaking 
area.  Using these assumptions, the ground test data can be extrapolated 
to the flight case and are plotted in figure 9.  Another curve, labeled 
35 K MLI Eccosorb, shows the effect of the same assumed incident 
radiation field, transmitted into the Eccosorb at its exposed rim, and
attenuated as it goes through.  This response turns sharply down at high
frequency because of the increasing Eccosorb absorption coefficient. 

The calibrator support arm can scatter radiation downward.  To estimate
this we assume that the sunshade has the same emissivity as the
insulation blankets, with two stages of geometric attenuation.  First,
the sunshade subtends only about 0.15 sr as seen from the calibrator
arm. Second, the radiation scattered by the arm edge is attenuated by a
factor of 0.05 to allow for the divergence of the scattered ray bundle
from the arm.  The net brightness is an order of magnitude less than the
emission from the multilayer insulation if it is at 35 K.  These 
assumptions produce a pair of curves similar to those for the multilayer 
insulation.

If any of the leakage sources were important in flight, our calibration 
with cold calibrator sources would have revealed an offset at high 
frequencies.  No such offset was observed.  The limit for this 
measurement is shown in figure 9.

\section{SUMMARY AND CONCLUSIONS}
 
The calibration methods for the Far Infrared Absolute Spectrophotometer (FIRAS) 
have been described and the accuracy estimated. The improvement in the
estimation of the uncertainty allows an improved absolute temperature measurement of
$2.725\pm 0.002$ K (95\% confidence). This estimate does not disagree with
earlier measurements which are less precise. The calibrator errors are all 
estimated to be smaller than the measured CMBR spectrum distortion limits 
reported by Mather \etal (1994), Fixsen \etal (1996) and Fixsen \etal (1998).  
 
\acknowledgements
 We thank A. Murdoch and H. Hemmati for their detailed measurements 
of the optical properties of Eccosorb and the reflectivities and 
transmissions of the calibrator assemblies, through several 
generations of designs.   W. Eichhorn made ray traces of the horn.
The calibrator was machined in the GSFC 
shops, and C. Clatterbuck developed the recipes for mixing and 
casting the Eccosorb blocks.

\clearpage
\typeout{FIGURE CAPTIONS}
\begin{figure}
\caption{Drawing of the FIRAS instrument. Light enters the Sky horn from the 
sky or the XCAL and the Reference horn from the ICAL. After reflection from
the folding flats (FL, FR) it bounces off the mirrors (ML, MR) and
is analyzed by the polarizer (A). The collimator mirrors (CL, CR) recollimate
the light before it is split by a second polarizer (B) at 45$^\circ$. It then
is reflected by the dihedral mirrors with different paths set by the 
mirror mechanism. After reflection the light retraverses the beamsplitter,
collimator mirrors and analyzer. This time it is intercepted by the 
pickoff mirrors (PL, PR) which direct it into the elliptical mirrors (EL, ER),
the dichroic filters and finally the detectors (Det\_LH, Det\_LL, Det\_RH, Det\_RL).}
\label{FIRAS}
\end{figure}

\begin{figure}
\caption{Cross section of the FIRAS calibrators. The XCAL is 140 mm in diameter
and the Ical is 60 mm in diameter.  Heaters and thermometers are indicated on 
drawing. The Hot Spot heater was designed to null a high frequency excess in
the CMBR. No excess was seen but the Hot Spot is part of the reason the 
ICAL has a reflectance of $\sim$ 4\%.}
\label{XCAL}
\end{figure}

\begin{figure}
\caption{Finite element model of calibrator. There are generally three layers:
the copper backing, the iron loaded epoxy, and the front surface. Each section 
is repeated for eight elements around the circumference as shown in the lower 
right. The additional thermometers on the ground test of the duplicate 
calibrator are indicated. There were other thermometers on the support arm
and the helium bath (not shown).  The multilayer insolation on the back
(upper surface) is also not shown.}
\label{final}
\end{figure}

\begin{figure}
\caption{Errors due to gradients. The possible errors due to a thermal gradient
from the surface to the inside (gradient), a spot 100 mK colder on the back 
of the Eccosorb, the limits from the measurements of y, and the measured 
temperature variation are shown.  A 0.1 mK temperature change and the FIRAS
uncertainies are shown for comparison. The curves with the dip at $\sim 8~\icm$
change sign there and are negative at lower frequency. The 2.7 K blackbody
peaks at 400 MJy/sr.}
\label{graderr}
\end{figure}

\begin{figure}
\caption{Schematic of the setup to measure the reflection of the 
calibrator.  From the left are the Magic Tee with four ports, connected to
(left) a load, (top) the detector, (front) an oscillator source, and (right)
a rectangular to cylindrical transition.  The circular waveguide has a conical
horn to match it to the throat of the horn.  The elliptical section of the
horn is not used.  The calibrator is moved near its nominal position (at far 
right) to measure the reflection. Measurements were done at 
33 and 94 GHz. See text for details.}
\label{reffreq}
\end{figure}

\begin{figure}
\caption{Plotted here are the measured reflectances as a function of tilt 
of the calibrator for 33 and 94 GHz. (See also fig 5.)}
\label{Refang}
\end{figure}

\begin{figure}
\caption{Shown here are plots of the reflectances versus frequency for several
modes of reflection. The X's with error bars are the measurements at 
33 and 94 GHz. Other lines are calculations; see text for details.}
\label{reffreq}
\end{figure}

\begin{figure}
\caption{Errors due to reflectances. The terms fall off at high frequency
because all of the emitters in the FIRAS instrument are below 3 K. The X's
with error bars are the measurements at 33 and 94 GHz.}
\label{referr}
\end{figure}

\begin{figure}
\caption{Errors due to leakage. The sunshade is assumed to heat the 
insolation layer to $\sim$35K.}
\label{leakerr}
\end{figure}

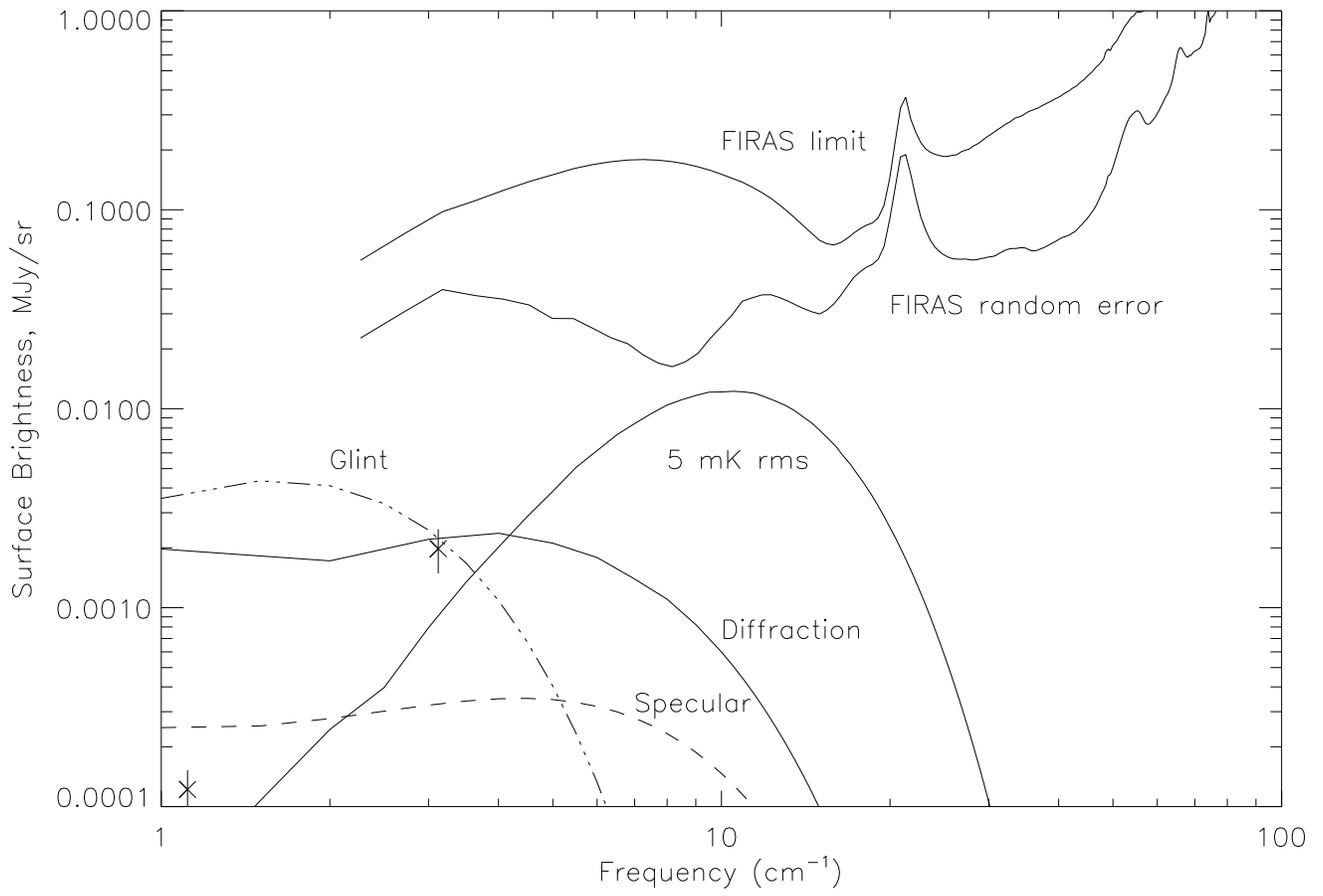
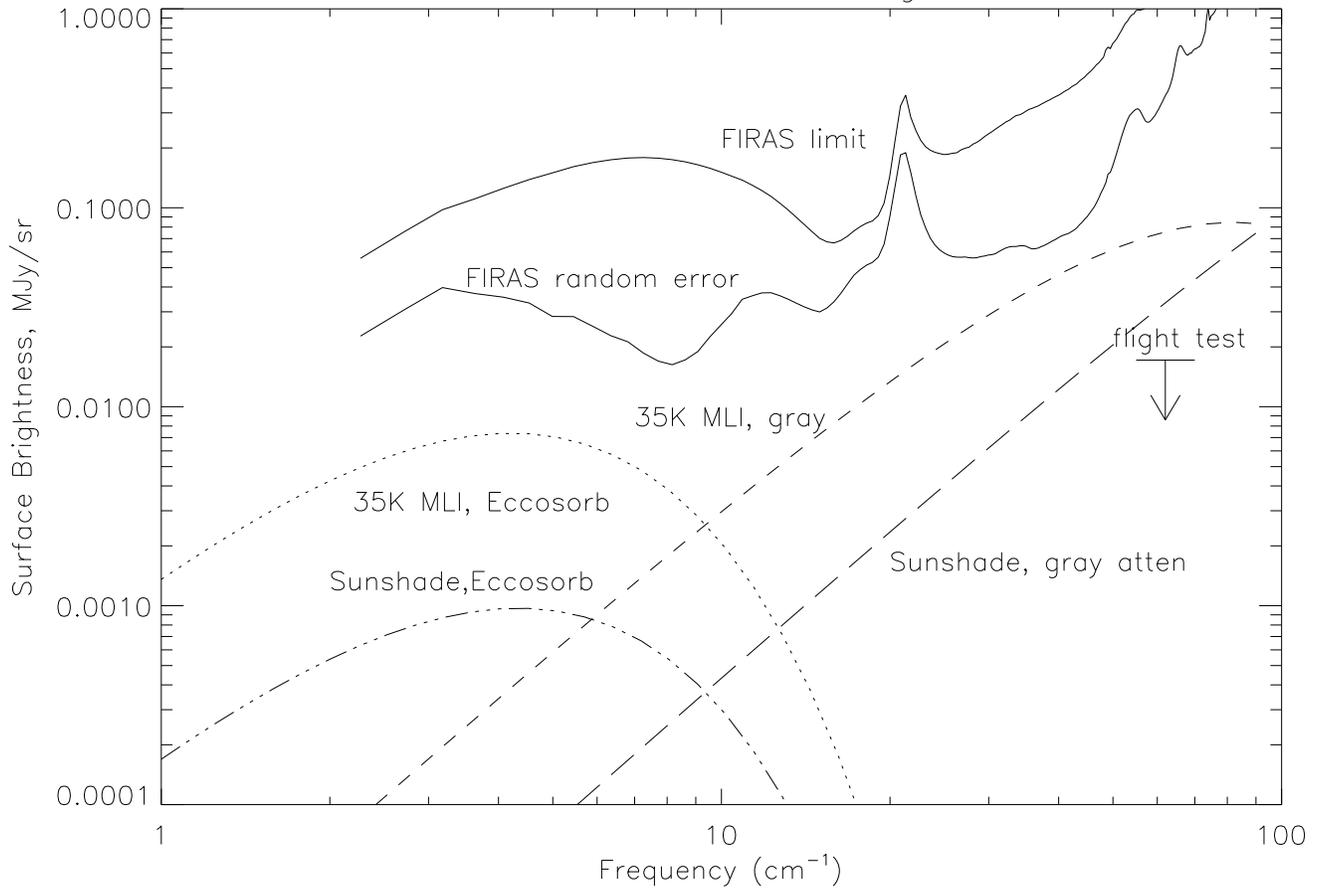
\end{document}